\documentclass[runningheads]{llncs}
\usepackage{graphicx}
\usepackage{booktabs}
\usepackage{hyperref}
\usepackage{listings}
\usepackage{xcolor}
\usepackage{todonotes}

\definecolor{rustkeyword}{RGB}{160,32,240}
\definecolor{rustcomment}{RGB}{0,128,0}
\definecolor{ruststring}{RGB}{163,21,21}

\lstdefinelanguage{Rust}{
  keywords={fn, pub, impl, unsafe, let, mut, as, self, use, mod, struct, enum, trait, where, for, in, match, if, else, loop, while, break, continue, return, crate, super},
  keywordstyle=\color{rustkeyword}\bfseries,
  ndkeywords={u8, usize, str, bool, Result, Option},
  ndkeywordstyle=\color{rustkeyword},
  comment=[l]{//},
  commentstyle=\color{rustcomment}\ttfamily,
  stringstyle=\color{ruststring}\ttfamily,
  morestring=[b]",
  sensitive=true
}

\begin{document}

\title{Verifying the Rust Standard Library}
\titlerunning{Verifying the Rust Standard Library}
\author{
Byron~Cook\textsuperscript{1,3} \and
Remi~Delmas\textsuperscript{1} \and
Zyad~Hassan\textsuperscript{1} \and
Bart~Jacobs\textsuperscript{4} \and
Ranjit~Jhala\textsuperscript{5} \and
Rahul~Kumar\textsuperscript{1} \and
Felipe~R.~Monteiro\textsuperscript{1} \and
Thanh~Nguyen\textsuperscript{1} \and
Rebecca~Rumbul\textsuperscript{2} \and
Michael~Tautschnig\textsuperscript{1,6} \and
Celina~Val\textsuperscript{1} \and
Carolyn~Zech\textsuperscript{7}
}

\authorrunning{Cook et al.}

\institute{
Amazon Web Services \and
Rust Foundation \and
University College London \and
KU Leuven \and
University of California, San Diego \and
Queen Mary University of London \and
Massachusetts Institute of Technology
}

\maketitle

\begin{abstract}
Rust's type system prevents many classes of memory errors, yet its
standard library relies heavily on \texttt{unsafe} code
whose correctness is validated through testing, including dynamic checks
under Miri, but lacks static verification.
We present what is, to the best of our knowledge,
the largest verification campaign reported for a
software library: an open, crowdsourced effort that integrates
complementary verification tools into the continuous
integration of a verification repository forked from the Rust
standard library.
We analyze the campaign's effectiveness, discuss the practical value
of machine-checked proofs for a subset of undefined behaviors (e.g., out-of-bounds access,
null and dangling pointer dereferences, and use of uninitialized memory), and
frame the remaining obstacles as open challenges for the
formal-methods community.
\end{abstract}

\section{Introduction}

%
Rust's ownership-based type system statically prevents data races and many
classes of memory errors, a guarantee that has driven adoption in operating
system kernels~\cite{Boos20}, browser engines~\cite{Anderson16}, cryptographic
libraries~\cite{Bhargavan25}, and safety-critical embedded systems~\cite{Levy17}.
Yet the guarantee has a gap.
The Rust standard library is widely regarded as battle-tested
software\footnote{\url{https://doc.rust-lang.org/std/}}, but it contains approximately
7,500 \texttt{unsafe} functions and 3,000 additional safe functions that
internally use \texttt{unsafe} code.
Inside these \texttt{unsafe} regions,
correctness relies on careful manual reasoning and testing rather than
on machine-checked proof~\cite{Astrauskas20,Cui24}.
Testing, including dynamic analysis under Miri~\cite{Jung26miri}, has
been effective at catching many issues, but it can only exercise a
finite set of executions.
Verification offers a qualitatively different assurance: a
machine-checked proof that a given property holds for \emph{all}
inputs, providing a permanent, auditable guarantee.

Landmark projects and industrial efforts have demonstrated that
verification can scale to compilers, microkernels, cryptographic
stacks, and continuous-integration
workflows~\cite{Leroy09,Klein09,Everest25,Jung18,Chong20,Chong21}.
However, a recent survey of 32 deployed verified
systems~\cite{Huang26} finds that even the largest efforts
verified codebases of at most hundreds of thousands of lines,
required dedicated teams of verification experts, and spanned
multiple person-years of effort.
The Rust standard library presents a qualitatively different challenge:
approximately 34{,}000~functions across the \texttt{core}, \texttt{alloc}, and \texttt{std} crates, a codebase
that ships a new release every six weeks, and a contributor base that spans
industry, academia, and the open-source community.
Verifying it demands not only advances in verification technology but also a
new model for organizing proof work at scale.

This paper reports on the first large-scale effort to prove the absence of
undefined behavior in the Rust standard library.
The project is organized as an open challenge program that invites contributions
from any individual or
team.
Since its launch in November~2024,
the effort has integrated four verification
tools (i.e., Kani~\cite{VanHattum22}, ESBMC~\cite{Gadelha18},
VeriFast~\cite{Jacobs11}, and Flux~\cite{Lehmann23}) into continuous
integration, with four additional tools under review.
An independent study of the project's early phase is reported by Le~Blanc and
Lam~\cite{LeBlanc25}.
To the best of our knowledge, and based on the most comprehensive
survey of deployed verified systems to date~\cite{Huang26}, this is
the largest verification campaign reported for a software
library, both by number of functions mechanically proved free of
classes of undefined behavior and by number of independent
contributors.

We make four contributions.
First, we describe the design of a crowdsourced verification campaign
organized around challenges with financial rewards, and analyze how this
structure enabled multi-tool integration and attracted contributions (Section~\ref{contest}).
Second, we present Autoharness, a tool that automatically generates
proof harnesses at the level of Rust's MIR intermediate representation;
Autoharness produced 16,748~harnesses, including 4,645 for unsafe functions
and 1,126 for safe abstractions over unsafe code, of
which 11,970 were verified against Kani's supported classes of undefined
behavior.
Across both automatic and manual harnesses, 989 functions were verified
against formal function contracts (Section~\ref{progress}).
Third, we report on the integration of four verification tools into
continuous integration, providing concrete evidence that multi-tool
verification of a production Rust codebase is practical
(Section~\ref{tools}).
Fourth, we identify and analyze the main technical
obstacles (e.g., generic function verification, intrinsic modeling, and
concurrency) and frame them as open problems for the formal-methods
community (Section~\ref{open-challenges}).
The project, its open challenges, and all verification artifacts are publicly
available\footnote{\url{https://github.com/model-checking/verify-rust-std}}.

\section{How is Rust unsafe?}
\label{motivation}

Rust's ownership-based type system statically prevents data races and many classes of
memory errors~\cite{Jung18}, but the standard library supplements these
guarantees with \texttt{unsafe} code that accesses five additional operations
the compiler does not check for memory safety: dereferencing raw pointers,
calling unsafe functions, accessing mutable statics, implementing unsafe traits,
and accessing fields of unions~\cite{Jung21}.
Importantly, \texttt{unsafe} Rust is still Rust, not C: the code is written
in the same language and compiled by the same compiler, but the programmer
takes responsibility for a small set of properties the compiler cannot verify.
The borrow checker, type checking, and lifetime analysis still apply inside
\texttt{unsafe} regions, but the soundness of these additional operations rests
on the developer's reasoning,
typically recorded only in natural-language
comments~\cite{Astrauskas20,Cui24}.
Over the last three years, developers have reported over 74 soundness
bugs\footnote{\url{https://github.com/rust-lang/rust/labels/I-unsound}, accessed March~2026};
18~CVEs have been filed
historically.\footnote{\url{https://rustsec.org/packages/std.html}, accessed March~2026}

Throughout this paper, \emph{absence of undefined behavior (UB)} refers to the
property that a function, when executed on any valid input, does not trigger
any of the behaviors that the Rust Reference classifies as
undefined~\cite{RustUBRef}.
Absence of undefined behavior is a sufficient condition for memory safety,
but not for full \emph{correctness}, which additionally
requires functional correctness, liveness, and security properties.
The current verification tooling does not check for all classes
of undefined behavior: Kani does not detect violations of the pointer aliasing
rules (as formalized by Stacked Borrows~\cite{Jung20} and Tree
Borrows~\cite{Villani25}), data races, invalid uses of inline assembly, or
all forms of provenance-related undefined behavior.
Section~\ref{progress} details the precise scope of each tool, and
Section~\ref{limitations} discusses the implications of this incomplete
coverage.

\section{Scaling the verification effort}
\label{contest}

The scale of the Rust standard library makes verification difficult to centralize
within a single team or institution.
To address this, we designed a community-oriented verification program, run by the
Rust Foundation\footnote{\url{https://rustfoundation.org/}}, that treats the
verification of the standard library as an open contest.

\subsection{Crowd-sourcing the verification effort}
\label{crowdsourcing}

Verification tasks are organized into \emph{challenges}.
Each challenge specifies a concrete verification target, a list of assumptions,
explicit success criteria, and a financial reward disbursed upon completion.
The status of all challenges is maintained in the project
website\footnote{\url{https://model-checking.github.io/verify-rust-std/}}.
Participants fork the repository, implement a candidate solution, and submit it as a
pull request reviewed by a technical review committee.
Accepted solutions are merged into a dedicated
fork\footnote{\url{https://github.com/model-checking/verify-rust-std}} of the Rust
repository that serves as the verification target.
The repository has received more than 450 pull requests from at least 21 unique
external contributors affiliated with four distinct institutions.
As of March 2026, the project has published 29 challenges.

Contributors specify contracts and loop invariants using the contract systems provided
by the participating tools.
For Kani-based Autoharness, these contracts express safety-related preconditions,
postconditions, type invariants, and loop conditions targeting absence of undefined
behavior.
The contest welcomes any verification tool for Rust programs, and some tools go
beyond safety properties: for example, the VeriFast proofs in
Section~\ref{sec:case-study} establish functional correctness of linked-list
operations.

\subsection{Tool Integration and Continuous Verification}
\label{tools}

The contest is designed to be tool-agnostic.
Any verification tool is eligible, provided it can operate on the Rust standard
library, can be integrated into continuous integration (CI), and provides clear
soundness guarantees, typically supported by peer-reviewed publications.
The project currently supports Kani~\footnote{\url{https://github.com/model-checking/
kani}}, ESBMC~\cite{Gadelha18},
Flux~\cite{Lehmann23}, and VeriFast~\cite{Jacobs11}.
Four additional tools are under review:
Verus~\cite{Lattuada24}, Creusot~\cite{Denis22}, KRust~\cite{Wang18}, and
RAPx~\footnote{\url{https://github.com/safer-rust/RAPx}}.
This diversity is essential because no single tool can discharge all verification
conditions; architecture-specific intrinsics, pointer-heavy code, concurrency
primitives, and loops with complex invariants require complementary reasoning
techniques.
Each tool targets a different set of properties:
Kani and ESBMC perform bounded model checking for memory safety violations
(e.g., out-of-bounds access, null and dangling pointer dereferences, use of
uninitialized memory, and arithmetic overflow), and support unbounded
analysis when loops and recursive functions are annotated with loop
contracts and function contracts, respectively;
Flux checks refinement types that encode numeric bounds and safety
preconditions;
and VeriFast uses separation logic to verify absence of all undefined
behavior, including pointer aliasing violations, for the functions it covers.

All verification is performed automatically through CI.
The verification repository is a maintained fork of upstream
\texttt{rust-lang/rust}, periodically synchronized so that proofs remain current.
Each pull request triggers the full suite of verification tools on all active proofs;
any violation is detected immediately
(cf.\ Section~\ref{findings}).
This continuous verification model, inspired by prior industrial
efforts~\cite{Chudnov18,Chong20,Chong21}, ensures that proofs remain valid across
revisions and guards against regressions.

\section{Verification progress}
\label{progress}

We automated the collection of both code and proof coverage metrics to track the state of the
verification effort over time.
Each CI run produces a JSON report that records, for every function in the standard library,
whether a proof harness exists, whether the harness succeeded, and which tool was responsible
for the proof.
As of the \texttt{nightly-2025-10-08} toolchain snapshot, the \texttt{core}, \texttt{alloc}, and \texttt{std} crates together
contain 33,955 functions.
The verification repository pins this toolchain version; all results below
reflect verification runs against this snapshot.
The metrics below cover all three crates.

\begin{figure}[!htbp]
\centering
\includegraphics[width=0.8\columnwidth]{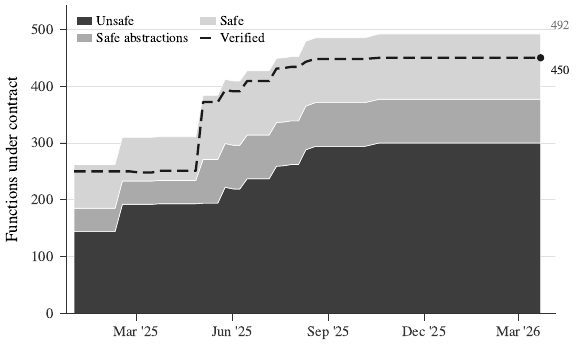}
\caption{Growth of manually written function contracts over time
(\texttt{core}~+~\texttt{std}), broken down by function category:
\emph{unsafe} functions (marked \texttt{unsafe fn}),
\emph{safe abstractions} (safe functions that internally use
\texttt{unsafe} blocks), and \emph{safe} (all other functions).
The dashed line shows the subset of contracts whose proofs pass
verification (always $\leq$ the total).
The plateau after October 2025 motivated the development of automatic
harness generation.}
\label{fig:contracts-over-time}
\end{figure}

Over the first sixteen months of the project, contributors manually wrote 725 proof
harnesses using Kani (694 with function contracts), plus more than 50 proofs constructed
with VeriFast.
Because the project began with manual proof engineering, these early harnesses span a wide range
of complexity, from straightforward type conversions to intricate pointer manipulations,
loop-heavy algorithms, and linked data structures requiring carefully crafted preconditions
and loop invariants.
Figure~\ref{fig:contracts-over-time} shows the growth of manually written
function contracts (i.e., precondition/postcondition annotations) and their
verified subset over time, broken down by function category.
After an initial ramp-up driven by the first wave of challenge solutions, the rate of new
contracts plateaued around October 2025.
At that point, manual harnesses covered only a small fraction of the standard library's
functions, leaving the vast majority unverified.
This plateau illustrates a fundamental scalability limitation: manual proof engineering alone
cannot reach the tens of thousands of functions in the standard library.

Even with limited coverage, the manual verification effort demonstrated how effective
function contracts could be for verifying the standard library, and this had a significant
institutional impact.
The Rust language team accepted contracts as an experimental language
feature\footnote{\url{https://doc.rust-lang.org/beta/unstable-book/language-features/contracts.html}},
and the Rust Project adopted a formal goal to instrument the standard library with safety
contracts\footnote{\url{https://rust-lang.github.io/rust-project-goals/2024h2/std-verification.html}}.

To overcome the scalability barrier, we developed \emph{Autoharness}, a tool that automatically
generates proof harnesses at the level of Rust's MIR intermediate representation.
A \emph{proof harness} is analogous to a unit test, but instead of calling a function
with concrete inputs, it calls the function with \emph{symbolic} (nondeterministic)
values that represent all possible inputs simultaneously.
A model checker then exhaustively explores all reachable execution paths,
turning the harness into a formal verification problem rather than a test.

\subsection{Autoharness: Design and Workflow}
\label{sec:autoharness}

Autoharness\footnote{\url{https://model-checking.github.io/kani/reference/experimental/autoharness.html}} is implemented as a compiler pass inside Kani.
Given a crate, it enumerates every function definition, determines which functions are
eligible for automatic harness generation, and for each eligible function synthesizes a
proof harness that calls the function on fully nondeterministic inputs.
The generated harness is then verified by Kani's back-end (e.g., CBMC), which
checks for the absence of its supported classes of undefined behavior (UB) along all reachable execution paths.

\paragraph{Eligibility filtering.}
A function is eligible if it satisfies three conditions:
(i)~it is \emph{monomorphic}, i.e., it has no unresolved generic type parameters;
(ii)~every argument type either implements or can automatically derive the
\texttt{kani::Arbitrary} trait, which generates a fully symbolic value
covering all bit-valid representations of that type; and
(iii)~it is not a Kani-internal implementation function.
Functions that fail any condition are recorded with a skip reason (e.g., generic, missing
\texttt{Arbitrary}, no body, or internal) and excluded from harness generation.
User-supplied include and exclude patterns can further restrict the set.

\paragraph{Nondeterministic inputs and type invariants.}
A key subtlety is the distinction between \emph{validity invariants} and
\emph{safety invariants}~\cite{RustUBRef}.
Validity invariants must hold at all times and are exploited by the compiler
for optimizations (e.g., a \texttt{bool} is always 0 or 1); Kani's
\texttt{Arbitrary} implementations respect these by construction.
Safety invariants, however, are semantic properties that safe code may assume
but that the type system does not enforce (e.g., \texttt{Duration}'s
nanosecond field must be less than $10^9$).
A derived \texttt{Arbitrary} implementation generates all bit-valid values
of a type but may violate its safety invariants, potentially causing
false positives when the function under test assumes those invariants hold.
To address this, we introduced an \texttt{Invariant} trait into the
standard library that allows types to express their safety invariants
programmatically via an \texttt{is\_safe()} predicate.
When a function carries a contract, Autoharness generates a
contract-verification harness that assumes the preconditions, which can
include \texttt{Invariant} checks, restricting inputs to states that
satisfy the type's safety invariants.
For functions without contracts, the proof is valid over all bit-valid
inputs; any failure indicates genuine UB, while the
absence of failure means the function is safe for all such inputs.
For unsafe functions specifically, this distinction is important.
An unsafe function without a contract is verified against all bit-valid
inputs with no assumed preconditions.
If the proof succeeds, the function cannot trigger any of Kani's
supported classes of UB regardless of how it is called,
a strong guarantee.
If it fails, the failure may reflect genuine UB or
a missing precondition that callers are expected to establish;
such functions require function/loop contracts to verify meaningfully.

\paragraph{Harness synthesis.}
For each eligible function~$f$ with argument types $\tau_1, \ldots, \tau_n$, Autoharness
constructs a new MIR function body that:
\begin{enumerate}
  \item allocates a local variable $x_i$ for each argument and initializes it by calling
  \texttt{kani::any::<$\tau_i$>()}, which produces a fully nondeterministic value of
  type~$\tau_i$;
  \item calls $f(x_1, \ldots, x_n)$.
\end{enumerate}
If $f$ carries a function contract (i.e., preconditions and postconditions),
the harness is generated as a contract-verification harness: Kani assumes
the preconditions, executes $f$, and checks the postconditions.
Similarly, if $f$ contains loop contracts, Autoharness detects and verifies them.
If $f$ has no contract, the harness simply calls $f$ on unconstrained inputs
(as shown in Listing~\ref{lst:autoharness-example}), and Kani
checks for its supported classes of undefined
behavior\footnote{\url{https://model-checking.github.io/kani/undefined-behaviour.html}}
during execution.
Crucially, all harnesses are generated internally at the MIR level without modifying the
crate's source code, which makes the approach suitable for continuous integration.

\begin{lstlisting}[language=Rust,caption={Generated harness for a safe wrapper (simplified). Kani verifies this harness by exploring all possible values of \texttt{self\_val} and \texttt{rhs}, checking that no execution path triggers UB. Because \texttt{wrapping\_shl} is defined to wrap on overflow, the proof succeeds for all inputs.},label={lst:autoharness-example}]
// Target function (in core::num)
pub const fn wrapping_shl(self, rhs: u32) -> u32 {
    unsafe { ... } // intrinsic call
}

// Harness generated by autoharness (conceptual)
#[kani::proof]
fn check_wrapping_shl() {
    let self_val: u32 = kani::any();
    let rhs: u32 = kani::any();
    let _ = self_val.wrapping_shl(rhs);
}
\end{lstlisting}

For argument types that do not have a source-level \texttt{Arbitrary} implementation,
Autoharness includes a companion pass (\texttt{AutomaticArbitraryPass}) that synthesizes
\texttt{Arbitrary} derivations for structs and enums at the MIR level, constructing each
field nondeterministically.
For enums, the pass generates a nondeterministic discriminant constrained to
the valid variant indices and initializes the fields of the selected variant;
variants with non-public fields or unsupported types cause the derivation to
fail, and the function is recorded as skipped.
The \texttt{Invariant} trait approach has a known limitation: types whose
safety invariants are not yet annotated with \texttt{is\_safe()} are tested
over all bit-valid values, which may include states that violate the type's
intended invariants.
This can produce false positives (e.g., harness failures on inputs that no
well-behaved caller would construct); in the current campaign, such
failures are classified as ``needs contract'' rather than as bugs.

\paragraph{Property discharged.}
Kani instruments the code with assertions that check for UBs including
out-of-bounds memory access, null and dangling pointer dereference, use of uninitialized
memory, and arithmetic overflow in unsafe contexts. For functions 
without loop contracts, Kani unrolls loops up to a configurable bound
and checks all reachable paths; an \emph{unwinding assertion} verifies
that the bound is sufficient, i.e., that no loop attempts to execute
beyond the unwinding limit, and reports a verification failure otherwise.
This yields a bounded proof of absence of the supported
classes of UB. When a function carries loop contracts,
Kani instead verifies that the invariant is inductive and uses it to abstract the loop,
replacing exhaustive unrolling with a single-iteration check from an arbitrary state satisfying the
invariant. A successful verification with loop contracts therefore constitutes an
inductive proof for the annotated loops, removing the dependency on the unwinding bound.
Kani does not yet support loop variant (\texttt{decreases}) clauses, so termination
is not verified; this is planned for future work.

\subsection{Manual vs.\ Automatic Harness Generation}
\label{sec:manual-vs-auto}

\begin{figure}[!htbp]
\centering
\includegraphics[width=0.8\columnwidth]{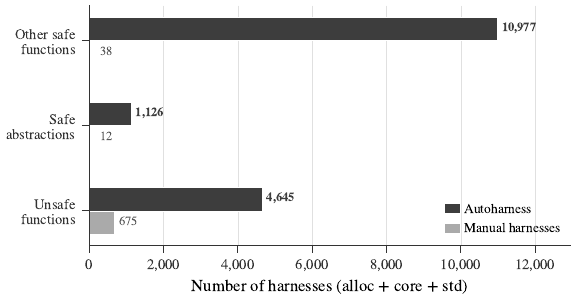}
\caption{Number of proof harnesses \emph{produced} (not necessarily verified)
by Autoharness vs.\ manual effort, across function categories
(\texttt{alloc}~+~\texttt{core}~+~\texttt{std}, March 2026).
Of the 16,748 automatically produced harnesses, 11,970 were successfully
verified (Table~\ref{tab:verified-breakdown}).}
\label{fig:coverage-comparison}
\end{figure}

Autoharness produced 16,748 proof harnesses (Figure~\ref{fig:coverage-comparison}),
including 4,645 for unsafe functions and 1,126 for safe abstractions, an
order-of-magnitude increase over the 725 manual harnesses accumulated over
sixteen months.
Not all produced harnesses pass verification: of the 16,748, 11,970 were
successfully verified against Kani's supported
classes of UB\footnote{Including out-of-bounds access, null and dangling pointer dereference,
use of uninitialized memory, and arithmetic overflow in unsafe contexts; see
\url{https://model-checking.github.io/kani/undefined-behaviour.html} for the full list.
Classes not yet covered include violations of the aliasing model and some forms of
provenance-related UB.};
the remaining 4,778 failed due to missing models, timeouts, or unsupported features.

\begin{table}[!htbp]
\centering
\footnotesize
\caption{Autoharness verification results by function category and
contract status (\texttt{alloc}~+~\texttt{core}~+~\texttt{std}).}
\label{tab:verified-breakdown}
\renewcommand{\arraystretch}{1.05}
\begin{tabular}{l r r r}
\toprule
\textbf{Category} & \textbf{With contracts} & \textbf{Without} & \textbf{Total verified} \\
\midrule
Unsafe functions        & 184  & 622    & 806   \\
Safe abstractions       & 44   & 926    & 970   \\
Safe functions          & 67   & 10,127 & 10,194 \\
\midrule
All                     & 295  & 11,675 & 11,970 \\
\bottomrule
\end{tabular}
\end{table}

\paragraph{Interpreting the verified count.}
Not all 11,970 verified functions carry the same evidential weight.
Table~\ref{tab:verified-breakdown} breaks down the Autoharness results by
function category and presence of contracts.
Including the 694 contract-verified manual harnesses, a total of
989 functions (295 automatic + 694 manual) were verified against formal
function contracts, the strongest guarantee the campaign provides.
The 10,194 verified safe functions (85\% of the Autoharness total)
provide unambiguous value: Kani symbolically executes through all
callees, including any unsafe code reached transitively, confirming
that no path triggers UB from Kani's checked classes on all bit-valid
inputs.
The 970 verified safe abstractions similarly exercise the unsafe code
they encapsulate.
For the 806 verified unsafe functions, the interpretation depends on
whether a contract is present.
Of these, 184 carry function contracts whose preconditions are assumed
during verification, yielding genuine safety proofs.
The remaining 622 pass on fully nondeterministic, bit-valid inputs
with no assumed preconditions.
A passing proof for such a function means either that it is trivially
safe for Kani's checked UB classes (and its \texttt{unsafe} marker
relates to properties Kani does not yet check, such as aliasing), or
that Kani's UB coverage is incomplete for the function's actual
behavior.
Distinguishing these two cases requires human review or complementary
analysis techniques.
These 622 functions are best understood as candidates surfaced by
automated triage, and they motivate both expanding Kani's UB coverage
and writing formal contracts for unsafe functions.

To complement Autoharness, we experimented with LLM-based contract
synthesis, translating natural-language \texttt{SAFETY} comments into
function contracts.
This approach is preliminary: the generated contracts require manual
review before merging, and a systematic evaluation of precision and
recall is left to future work.

Despite its effectiveness, Kani cannot yet handle all functions.
Of the 17,207 functions skipped during automatic harness generation,
the dominant reason is generic type parameters (9,635 functions, 56\%):
Rust monomorphizes generics at compile time, and Autoharness currently cannot synthesize
inputs for uninstantiated type parameters.
The second largest category is missing \texttt{Arbitrary} trait implementations in Kani
for the argument types of the target functions (3,826 functions, 22\%), which prevents the
generation of nondeterministic inputs for types such as raw pointers, references, and complex
standard library types.
Internal Kani implementation functions account for a further 3,246 (19\%).
The remaining 500 functions (3\%) were skipped for miscellaneous reasons
including foreign function declarations and unsupported argument types such
as function pointers.
These limitations represent concrete targets for future tool development:
extending Kani's \texttt{Arbitrary} support to cover pointer and reference
types, and supporting generic function verification, would together address
78\% of the currently skipped functions.

The rapid growth in harness count created a new engineering challenge: compilation time.
Kani compiles Rust code into a GOTO-style intermediate representation before symbolically
executing it, and the sheer volume of 16,748 harnesses made this pipeline a bottleneck.
To address this, we developed dedicated compiler benchmarking tools to expose performance
bottlenecks and detect regressions in pull requests.
Guided by these benchmarks, we introduced parallelization, caching, heuristically ordered
code generation, and function stubbing, which together produced a $3.97\times$ speedup in
compilation for the standard library.
This improvement made it practical to run thousands of proofs in CI within acceptable time budgets.

\subsection{Case Study: Verifying \texttt{LinkedList} with VeriFast}
\label{sec:case-study}

The results presented so far are dominated by model checking via Autoharness.
The challenge system, however, also produced proofs whose
guarantees are qualitatively different: they reason about unbounded inputs
and establish that verified functions cannot cause UB for
any well-typed caller.
We describe one such proof to make this contrast concrete.

The VeriFast proof of \texttt{LinkedList}\footnote{\url{https://github.com/model-checking/verify-rust-std/pull/238}}
targets one of the most pointer-intensive modules in
the standard library: every public operation manipulates raw
\texttt{NonNull} pointers, manually manages heap allocations, and must
preserve a cycle-free invariant across insertions, removals, splits, and
cursor traversals.
The proof directly verifies 19 functions (e.g., \texttt{push\_front},
\texttt{split\_off}, \texttt{remove\_current}) and implies the soundness
of 5~additional non-\texttt{unsafe} functions (e.g., \texttt{contains},
\texttt{remove}, \texttt{drop}) that call only the verified functions.
For each function, the proof establishes \emph{soundness}: the function
will not exhibit UB when called by a well-typed caller.

\begin{lstlisting}[language=Rust,caption={Separation-logic predicate \texttt{Nodes} and contract for \texttt{push\_front\_node}. The predicate recursively asserts exclusive ownership of each heap-allocated node. The contract transfers ownership of the new node into the list.},label={lst:verifast-nodes}]
/*@ pred Nodes<T>(alloc_id: alloc_id_t, n: Option<NonNull<Node<T>>>,
                  prev: Option<NonNull<Node<T>>>,
                  last: Option<NonNull<Node<T>>>,
                  next: Option<NonNull<Node<T>>>;
                  nodes: list<NonNull<Node<T>>>) =
    if n == next {
        nodes == [] &*& last == prev
    } else {
        n == Option::Some(?n_) &*&
        alloc_block_in(alloc_id, n_.as_ptr() as *u8,
        Layout::new::<Node<T>>()) &*&
        struct_Node_padding(n_.as_ptr()) &*&
        (*n_.as_ptr()).prev |-> prev &*&
        (*n_.as_ptr()).next |-> ?next0 &*&
        pointer_within_limits(&(*n_.as_ptr()).element) == true &*&
        Nodes(alloc_id, next0, n, last, next, ?nodes0) &*&
        nodes == cons(n_, nodes0)
    }; @*/

unsafe fn push_front_node(&mut self, node: NonNull<Node<T>>)
    /*@
    req thread_token(?t) &*& *self |-> ?self0 &*&
        Allocator(t, self0.alloc, ?alloc_id) &*&
        Nodes(alloc_id, self0.head, None, self0.tail, None, ?nodes) &*&
        length(nodes) == self0.len &*&
        *node.as_ptr() |-> ?n &*&
        alloc_block_in(alloc_id, node.as_ptr() as *u8,
                       Layout::new::<Node<T>>());
    @*/
    /*@
    ens thread_token(t) &*& *self |-> ?self1 &*&
         Allocator(t, self1.alloc, alloc_id) &*&
         Nodes(alloc_id, self1.head, None, self1.tail, None,
               cons(node, nodes)) &*&
        self1.len == 1 + length(nodes) &*&
        (*node.as_ptr()).element |-> n.element;
    @*/
\end{lstlisting}

Central to the proof is a separation-logic predicate \texttt{Nodes}
(see Listing~\ref{lst:verifast-nodes}) that recursively asserts exclusive
ownership of each heap-allocated node between the list's \texttt{head} and
\texttt{tail} pointers.
Because the predicate's clauses are joined by a separating conjunction
(\texttt{\&*\&}), the predicate structurally rules out aliasing and cycles.
Each function receives a \texttt{req}/\texttt{ens} contract expressed in
terms of \texttt{Nodes}; Listing~\ref{lst:verifast-nodes} shows the
contract for \texttt{push\_front\_node}, where the precondition requires
separate ownership of the new node and the existing list, and the
postcondition returns a list whose node sequence is extended by one element.
VeriFast then symbolically executes the function body, checking that every
raw-pointer dereference, field mutation, and \texttt{Box::from\_raw\_in}
call respects the ownership discipline.

For safe functions such as \texttt{clear} and
\texttt{len}, VeriFast additionally verifies \emph{semantic
well-typedness}: the specification implies that the function cannot cause
UB for any well-typed caller.
The proof currently assumes that the allocator's lifetime equals
\texttt{'static} (i.e., the global allocator).
VeriFast does not yet fully verify Rust's pointer aliasing rules as
formalized by Tree Borrows~\cite{Villani25}: it verifies immutability of
shared references while in use, but does not check the aliasing
restrictions for mutable references or boxes, nor that references remain
valid for the duration of a function call.
This is a known soundness limitation, meaning the proof could miss
aliasing-related UB.

A distinctive feature of this proof is its treatment of source
modifications.
VeriFast requires minor code changes to insert ghost commands (e.g.,
replacing \texttt{for} loops with \texttt{loop} loops, or replacing
\texttt{Option::map} with first-order equivalents).
To ensure these changes do not alter program behavior, the proof includes a
\emph{refinement checking step} that mechanically verifies that every behavior of
the original code is also a behavior of the annotated version.
The CI pipeline runs three steps in sequence: VeriFast checks the annotated
code, the refinement checker relates it to the original, and a diff ensures
the original matches the upstream repository.
This three-stage pipeline illustrates the engineering overhead that
deductive verification tools impose relative to model
checking, but also the deeper guarantees they provide.

\subsection{Bugs Found and Fixed}
\label{findings}

The verification effort has not uncovered any previously unknown memory safety
vulnerabilities in the standard library.
This null result is itself informative: it speaks to the effectiveness of
Rust's existing testing infrastructure and Miri-based dynamic
analysis~\cite{Jung26miri} at catching memory safety bugs before they
reach production.
The primary value of the verification campaign is therefore not
bug-finding but the guarantee it provides: a machine-checked proof,
for each verified function, that the targeted classes of undefined
behavior cannot occur.
Testing can demonstrate the absence of bugs on exercised inputs;
verification certifies their absence on all inputs within scope.

The effort has, however, revealed concrete specification and documentation issues,
summarized in Table~\ref{tab:bugs}: missing safety annotations, incorrect
\texttt{SAFETY} comments, and documentation errors that misrepresented function
behavior.
These findings illustrate a secondary benefit: the process of writing formal
specifications forces a precise articulation of safety requirements that
natural-language comments alone do not provide.

\begin{table}[!htbp]
\centering
\caption{Issues found through the verification effort.}
\label{tab:bugs}
\renewcommand{\arraystretch}{1.1}
\begin{tabular}{l l l l}
\toprule
\textbf{Issue type} & \textbf{Component} & \textbf{Found via} & \textbf{Status} \\
\midrule
Incorrect SIMD shift results  & \texttt{stdarch}  & Challenge~15 & Fixed upstream \\
Missing \texttt{unsafe} annotations & \texttt{core} & VeriFast proofs & Fixed upstream \\
Incorrect \texttt{SAFETY} comments  & \texttt{core} & Flux proofs    & Fixed upstream \\
Incorrect panic documentation       & \texttt{core} & Kani proofs    & Fixed upstream \\
\bottomrule
\end{tabular}
\end{table}

\section{Lessons learned}
\label{lessons}

We reflect on three retrospective insights from the project: the cost of
community coordination, specification design for unsafe code, and the need
for tool diversity.

\subsection{Planning for Consensus and Community}

Our initial planning focused on technical milestones, but we significantly
under-budgeted time for activities that depended on community consensus and
institutional coordination.
Integrating function contracts into the Rust compiler required broad support
from the language team, library maintainers, and tool authors; even
technically straightforward changes stalled when stakeholders had not been
engaged early.
Attracting external contributors required not only financially rewarded
challenges but also a spectrum of task difficulty, clear documentation, and
worked examples.
Coordinating with external institutions required formal legal agreements that
took months to finalize.
The lesson is that upstream language changes, contributor onboarding, and
institutional agreements should be treated as first-class milestones with
explicit timelines, not as incidental tasks.

\subsection{Specification Design for Unsafe Code}

One of the more subtle lessons concerns the design of specifications in the
presence of immediate undefined behavior.
Rust defines\footnote{\url{https://doc.rust-lang.org/reference/behavior-considered-undefined.html}}
certain violations as triggering undefined behavior at the point
where an invalid value or reference is created, not when it is first used.
This has implications for where and how safety properties can be expressed.
Le Blanc and Lam~\cite{LeBlanc25} independently identified this challenge;
our experience confirms and extends their observation.

Consider \texttt{slice::from\_raw\_parts}, which takes a pointer and a length
and returns a slice reference.
If the function constructs a misaligned reference, undefined behavior occurs
immediately, before any postcondition can be checked.
In simple cases one can move the property into a precondition on the inputs,
but for functions where the source of undefined behavior arises deeper in the
call stack or depends on intermediate computations, expressing the right
property as a precondition becomes much harder.
We encountered this pattern in pointer-arithmetic functions, slice
constructors, and transmute wrappers throughout \texttt{core}.
The lesson is that specification languages aimed at safety verification need
a principled way to express \emph{internal} safety properties that must hold
at particular program points, not only at function boundaries.


\subsection{Tool Diversity in Practice}

The initial plan relied almost exclusively on Kani, but the project evolved
to include ESBMC, VeriFast, and Flux, driven by concrete technical
needs.
As discussed in Section~\ref{tools}, no single tool suffices: linked data structures require
unbounded separation-logic reasoning, while model checking is more
convenient for non-heap-intensive code.
Beyond technical complementarity, the multi-tool design gives the Rust
community comparative evidence on which tools and proof strategies justify
their maintenance cost in a continuously evolving codebase.

This diversity has implications for the Rust contract language initiative.
The experimental contract syntax adopted by the Rust language provides a
shared baseline for expressing preconditions and postconditions as boolean
expressions, and tools such as Kani and ESBMC can consume these directly.
However, richer verification approaches require specification constructs
that fall outside this baseline: VeriFast relies on separation-logic
predicates and ghost state to reason about heap ownership, while Flux uses
refinement types to track value-level invariants.
Our experience suggests that the Rust ecosystem will need to accommodate
tool-specific annotation layers alongside the common contract syntax,
rather than converging on a single specification language.

\section{Open technical challenges}
\label{open-challenges}

Despite verifying over 10{,}000 functions, the Rust standard library remains far from
fully verified.
Many high-impact APIs, including \texttt{BTreeMap} internals, atomic types, \texttt{String},
iterators, vectors, deques, and reference-counted types, are only partially covered or not
yet addressed.

\subsection{Intrinsics and Model Coverage}
\label{sec:intrinsics}

Of the 4,778 harnesses that fail under the standard CI settings with Autoharness, a
significant subset traces back to missing models: 71 unsupported Rust intrinsics and
813 unmodeled library functions, 721 of which are LLVM-internal SIMD intrinsics.
The Rust standard library relies heavily on these compiler intrinsics and foreign
functions whose semantics are defined outside of Rust itself.
Any verification tool must either model these operations faithfully
or treat them as opaque, which blocks verification of all callers.

This is not a limitation of any single tool.
Bounded model checkers, deductive verifiers, and abstract interpreters alike need
accurate models of intrinsics to reason about the code that calls them.
Closing this gap requires a shared, tool-agnostic effort: developing formal
specifications for the most frequently used intrinsics that any verification backend
can consume.
For SIMD intrinsics in particular, the randomized testing approach used in Challenge~15
(cf.\ Section~\ref{tools}) may serve as a pragmatic intermediate step, providing high
confidence where full formal models are not yet available.
A systematic triage process to distinguish model gaps from genuine specification or
code issues would further accelerate progress across all integrated tools.

\subsection{History-Dependent and Parametric Specifications}

Among failures not attributable to missing models, two recurring patterns
stand out.

\paragraph{History-dependent properties.}
Some functions require that a value has a particular provenance or
initialization history: a \texttt{MaybeUninit<T>} must be fully
initialized, or a pointer must originate from a specific constructor
(e.g., \texttt{Rc::from\_raw} expects its argument was produced by
\texttt{Rc::into\_raw}).
Expressing such properties as boundary-level contracts is often
insufficient (cf.\ Section~\ref{lessons}); they require tracking state
at internal program points.
Shadow memory instrumentation (Kani's \texttt{uninit-checks}),
separation-logic ghost state (VeriFast), and abstract interpretation
each address parts of this problem; extending them to cover unions
and provenance tracking is future work.

\paragraph{Generic functions.}
Generic functions represent the single largest category of unverified code:
automatic harness generation currently skips 9,635 of them because Rust
monomorphizes generics at compile time.
Bounded model checkers can enumerate representative type instantiations,
but this may miss relevant cases and scales poorly.
Deductive verifiers and refinement type systems can reason parametrically,
verifying a generic body once for all instantiations, at the cost of
richer specifications.
A hybrid strategy that routes type-agnostic functions to instantiation-based
checking and layout-dependent functions to parametric reasoning is a key
open problem.

\subsection{Concurrency and Data Races}

Two of the 29 challenges target concurrency-related APIs (i.e., Challenge~7: atomic types
and Challenge~27: \texttt{Arc}), yet neither has received accepted solutions.
Data races in unsafe Rust constitute immediate undefined behavior, making this a
significant gap in coverage.

VeriFast's separation-logic foundation inherently verifies absence of data races
and can check the proof obligations implied by \texttt{Send} and \texttt{Sync}
implementations; its test suite already includes proofs of simplified \texttt{Mutex}
and \texttt{Arc} implementations using sequentially consistent atomics.
However, the core difficulty for the standard library is specifying and verifying
lock-free data structures under \emph{relaxed} memory~\cite{Dang20,Jacobs25arc}.
Types such as \texttt{Arc} and the atomic primitives rely on ordering modes
ranging from relaxed to sequentially consistent, whose correctness depends
on subtle inter-thread invariants that require a formal memory model to express.
Tools such as VeriFast~\cite{Jacobs11}, Verus~\cite{Lattuada24}, and
Gillian-Rust~\cite{Ayoun25} target multi-threaded Rust, but applying their
concurrency reasoning to the standard library's lock-free types is future work.

\section{Limitations and threats to validity}
\label{limitations}

We consolidate the main limitations of the verification results reported in this paper.

\paragraph{Bounded reasoning.}
Kani, the primary verification backend for Autoharness, performs bounded model checking
by default: loops are unrolled up to a configurable bound.
Kani inserts unwinding assertions that detect when the bound is insufficient;
if these assertions pass, the proof covers all reachable paths and is complete
for the given function.
If the bound is too low, the unwinding assertion fails and the proof is reported
as unsuccessful.
In practice, proof harnesses may use \texttt{kani::assume} to restrict the size
of inputs so that unwinding assertions pass; such assumptions are explicit in the
harness and limit the proof to the assumed input range.
Kani also supports loop contracts, which replace exhaustive unrolling with an
inductive argument and yield proofs that do not depend on the unwinding bound.
Termination is not yet verified, as Kani does not yet support loop variant
(\texttt{decreases}) clauses.
Loop contracts must currently be supplied either manually or through the LLM-based
contract synthesis approach described in Section~\ref{sec:manual-vs-auto}.

\paragraph{Partial property coverage.}
Kani checks a subset of the undefined behaviors listed in the Rust
Reference~\cite{RustUBRef}.
Notably, it does not detect violations of the pointer aliasing model
(Tree Borrows~\cite{Villani25}/Stacked Borrows~\cite{Jung20}, both
implemented in Miri~\cite{Jung26miri}), data races, invalid inline
assembly, or all forms of provenance-related UB.
Nor does it verify that type safety invariants are preserved across
unsafe boundaries (e.g., that a \texttt{Vec}'s length does not exceed
its capacity).
A function that passes all current checks may still exhibit undefined
behavior under a more complete model.

\paragraph{Specification and model gaps.}
Rust does not yet have a ratified formal specification; key aspects such
as the aliasing model have competing proposals~\cite{Jung20,Villani25},
and a future specification change could invalidate proofs that are sound
under today's assumptions.
At the implementation level, verification of functions that call compiler
intrinsics or foreign functions depends on the fidelity of the models
provided for those operations (cf.\ Section~\ref{sec:intrinsics}).
Where models do exist, any inaccuracy can lead to unsound verification results.

\paragraph{Randomized testing for SIMD intrinsics.}
Challenge~15 was completed using randomized testing of executable models for 565 SIMD
intrinsics, not formal proof.
While the testing used logged seeds for reproducibility and uncovered two upstream bugs,
it does not provide the same guarantee as model checking or deductive reasoning.
These results should be interpreted as high-confidence validation rather than formal
verification.

\paragraph{Tool-chain soundness.}
All verification results depend on the correctness of the underlying tool
chains, including Kani's MIR-to-GOTO translation, CBMC's bounded model
checker, the SAT/SMT solvers invoked during verification, and VeriFast's
symbolic execution engine.
A soundness bug in any of these components could cause a genuine
undefined behavior to be missed.
None of these tools have been formally verified themselves; their
trustworthiness rests on extensive testing, fuzzing, and years of use in
production settings.

\paragraph{Harness-level vs.\ function-level claims.}
Each Autoharness proof verifies a single function in isolation, called on fully
nondeterministic inputs.
This does not account for calling-context constraints: a function that is UB-free on
all inputs is also UB-free in any calling context, but a function that relies on
preconditions established by its callers may fail the harness even though it is safe in
practice.
Conversely, when an unsafe function's preconditions are specified only in
natural-language comments, the proof does not verify that callers satisfy them.
However, when preconditions are expressed as \texttt{\#[kani::requires]} contracts,
Kani does check them at each call site: a harness for a caller will fail if the
caller does not satisfy the callee's contract.
Compositional reasoning, where caller proofs discharge callee preconditions, is
supported through manual function contracts but is not yet automated.

\section{Related work}
\label{related}

Landmark verification projects have delivered strong guarantees for
compilers~\cite{Leroy09}, microkernels~\cite{Klein09},
file systems~\cite{Chen15}, concurrent OS kernels~\cite{Gu16},
full software stacks~\cite{Hawblitzel14}, and cryptographic
libraries~\cite{Everest25}.
A survey of 32~deployed verified systems~\cite{Huang26} confirms that
these efforts verify single, fixed-version artifacts in depth, carried
out by dedicated teams over multiple years.
Our work targets a different point in the design space: broad coverage
of an evolving, multi-component library through a community-driven,
multi-tool campaign.

Within the Rust ecosystem, RustBelt provides a semantic foundation for
the type system and unsafe abstractions~\cite{Jung18}, and the aliasing
discipline is formalized by Stacked Borrows~\cite{Jung20} and its
successor Tree Borrows~\cite{Villani25}.
Empirical studies characterize the prevalence and documentation
challenges of unsafe code~\cite{Astrauskas20,Cui24}.
Verification tools with complementary strengths include
Prusti~\cite{Astrauskas22}, Creusot~\cite{Denis22},
Flux~\cite{Lehmann23}, Verus~\cite{Lattuada24},
Gillian-Rust~\cite{Ayoun25}, RefinedRust~\cite{Gaher24}, and
hax~\cite{Bhargavan25hax}.
Rather than proposing a new tool, we study how multiple existing tools
can be combined to verify a large, evolving library.

Complementary to static verification, the Miri interpreter~\cite{Jung26miri}
dynamically detects undefined behavior, including aliasing violations
under Stacked Borrows and Tree Borrows, use of
uninitialized memory, and data races.
Miri operates on concrete inputs and supports a broader subset of Rust
than Kani, including generic code and concurrency primitives, but does
not support foreign functions.
Miri has found dozens of bugs in the standard library and is integrated
into its CI; integrating it into the verification workflow for functions
beyond model checking's reach is a promising direction.

The closest line of work comes from continuous formal verification in
industrial settings: prior efforts have integrated model checking and
continuous assurance into CI workflows~\cite{Chong20,Chong21,cook21,Chudnov18},
and AutoCorres~\cite{Greenaway14} automatically abstracts C code into
higher-level proof representations.
Our work extends this perspective to a community-driven setting,
combining crowdsourced verification with automated harness generation
and multi-tool integration.
Le~Blanc and Lam~\cite{LeBlanc25} independently study the early phase
of this project; our paper covers the full campaign through March~2026,
including automated harness generation and quantitative results across
all three crates.

\section{Conclusion}

This paper reports on the first large-scale, community-driven effort to
verify the Rust standard library.
The campaign integrated four verification tools into continuous integration,
produced 16,748 automatic proof harnesses of which 11,970 were verified
against Kani's supported classes of undefined behavior, and established
989 contract-verified proofs across automatic and manual harnesses.
Along the way, the effort uncovered specification inconsistencies, missing
safety annotations, and documentation errors, and contributed to the
adoption of function contracts as an experimental Rust language feature.

Significant challenges remain: verifying generic functions, modeling compiler
intrinsics, reasoning about concurrency under relaxed memory, and keeping
proofs synchronized with an upstream repository that ships a new release
every six weeks; scaling this process will require upstreaming contracts and
proofs so that proof maintenance becomes part of the standard development
workflow.
These are problems where advances in deductive verification, abstract
interpretation, type-theoretic reasoning, and dataflow analysis could each
make a substantial contribution.
Our experience also suggests that the Rust ecosystem will need a layered
specification approach, with a shared contract syntax complemented by
tool-specific annotation layers, rather than a single universal
specification language.

Rust is now deployed in operating system kernels, browser engines,
cryptographic libraries, and safety-critical embedded systems.
By verifying foundational libraries like the standard library, we can help
ensure that developers are building on solid foundations.
The tools, infrastructure, and community assembled by this project provide
a concrete starting point toward that goal.

\begin{credits}
\subsubsection{\ackname}
We thank the Rust Foundation and all open-source contributors for their support.
An AI assistant was used during the preparation of this
manuscript for prose editing and bibliographic verification.
All technical content, experimental data, and scientific claims
are the sole work of the authors, who reviewed and approved the
final text.
\end{credits}

\bibliographystyle{splncs04}
\bibliography{refs}

\end{document}